\documentclass[aps, prl, showpacs, superscriptaddress, twocolumn]{revtex4}

\usepackage[utf8]{inputenc}
\usepackage{siunitx}
\usepackage{graphicx}
\usepackage{amsmath,amssymb}

\usepackage{xcolor}

\begin{document}
\title{Fracture Diodes: Directional asymmetry of fracture toughness}

\author{N. R. Brodnik$^*$}
\affiliation{Department of Mechanical Engineering, University of California, Santa Barbara, CA 93106, USA}

\author{S. Brach$^*$}
\affiliation{Laboratoire de M\'ecanique des Solides, \'Ecole Polytechnique, 91120 Palaiseau, France}

\author{C. M. Long}
\affiliation{Division of Engineering and Applied Science, California Institute of Technology, Pasadena, CA 91125, USA}

\author{G. Ravichandran}
\affiliation{Division of Engineering and Applied Science, California Institute of Technology, Pasadena, CA 91125, USA}

\author{B. Bourdin}
\affiliation{Department of Mathematics, Department of Mechanical \& Industrial Engineering, Louisiana State University, Baton Rouge, LA 70803, USA}

\author{K. T. Faber$^\dagger$}
\affiliation{Division of Engineering and Applied Science, California Institute of Technology, Pasadena, CA 91125, USA}

\author{K. Bhattacharya$^\dagger$}
\affiliation{Division of Engineering and Applied Science, California Institute of Technology, Pasadena, CA 91125, USA}

\begin{abstract}
 Toughness describes the ability of a material to resist fracture or crack propagation.  
It is demonstrated here that fracture toughness of a material can be asymmetric, i.e., the resistance of a medium to a crack propagating from right to left can be significantly different from that to a crack propagating from left to right.  
Such asymmetry is unknown in natural materials, but we show that it can be built into artificial materials through the proper control of microstructure.  
This paves the way for control of crack paths and direction, where fracture -- when unavoidable -- can be guided through pre-designed paths to minimize loss of critical components.  
\end{abstract}


\maketitle

\def\thefootnote{*}\footnotetext{These authors contributed equally to this work}\def\thefootnote{\arabic{footnote}}
\def\thefootnote{$\dagger$}\footnotetext{Corresponding authors: ktfaber@caltech.edu, bhatta@caltech.edu}\def\thefootnote{\arabic{footnote}}

It is not uncommon for a material system to exhibit anisotropy or orientation dependence in its mechanical properties.  
It can arise from the anisotropy of electronic interaction and atomic arrangement, as in the elastic moduli and fracture toughness of crystalline solids.  
It can also arise from the anisotropy of the heterogeneous structure as in both natural (e.g. sea shells, wood) and engineered (e.g. fiber-reinforced composite) materials.  A straightforward example is layered composite systems:  here, both elastic stiffness and failure strength can be drastically different depending on whether or not the direction of loading is into or out of the plane of lamination.  

While anisotropy is common, it is generally centro-symmetric; the property is invariant with reversal of direction.  
However, recent work has provided examples of interfacial phenomena where this symmetry is broken.  
Inspired by nature where textured surfaces enable butterflies to shed water from their wings, water striders to glide on water and plants to collect water, various researchers have developed gradient surfaces (periodic channels with increasing width~\cite{detal_lang_04,kt_apl_09}), textured surfaces (with pillars of increasing spacing~\cite{scb_lang_06} or with  asymmetric sawtooth patterns~\cite{setal_pre_99}) or surfaces with a unidirectionally slanted nano-rod array~\cite{metal_natmat_10} to transport droplets preferentially in one direction~(\cite{hsd_afm_19} for a recent review).  
Textured surfaces have been used in tribology for directional friction coefficients~\cite{letal_triblet_19}.   
Similarly, it is recently shown that adhesion can be direction specific~(\cite{xetal_prl_12,xetal_jmps_13,xetal_jmps_15} in adhesive tapes and~\cite{metal_sm_12} using subsurface liquid filled microchannels).  
However, all of these works concern interfacial phenomena.  

In this letter, we show that directional asymmetry extends to bulk phenomena, and in particular, to fracture. This was suggested in numerical simulations of Hossain {\it et al.}~\cite{Hossain-2014}.   We do so in the context of composite or metamaterials where the scale of the microstructure is small compared to the scale of the application.  The advent of additive manufacturing and 3D printing has enabled fine control of material structure giving rise to what is now often referred to as `metamaterials'.   
This precise control of microstructure has been exploited to develop metamaterials with unusual mechanical properties including those with chiral character~\cite{fkw_science_17} or  topologically protected modes~\cite{letal_repprogphys_15} (see~\cite{ketal_natrev_19} for a comprehensive review of 3D metamaterials).  
However, these concern deformation modes and wave propagation, and the study of failure is limited.  

Failure of a heterogeneous medium like a metamaterial is a complicated process.
At the microscale, the  stress is not uniform, and so the driving force on a crack tip depends on position, as does the resistance to crack growth (toughness).  Interfaces may pin or deflect cracks, and daughter cracks can nucleate distally.  The  stress at any point in time depends on prior history or prior crack trajectory.  So, the fracture process is neither uniform nor steady, one can have microscopic damage without macroscopic failure, and a sufficiently large macroscopic driving force  is necessary for the fracture process to progress at a macroscopic scale.   {\it We define the effective toughness (effective energy release rate) as the smallest driving force (energy release rate) necessary at the macroscale to drive the fracture process on the macroscale.}  Unlike elastic moduli and plastic strength, the effective toughness can be larger than those of the constituent materials, and has been exploited to toughen ceramics \cite{fe_acta} and composites \cite{hh_ijss_89}.  This letter shows that microstructure can lead to unexpected fracture properties like asymmetry.

\begin{figure}
	\centering
	\includegraphics[width=.95\columnwidth]{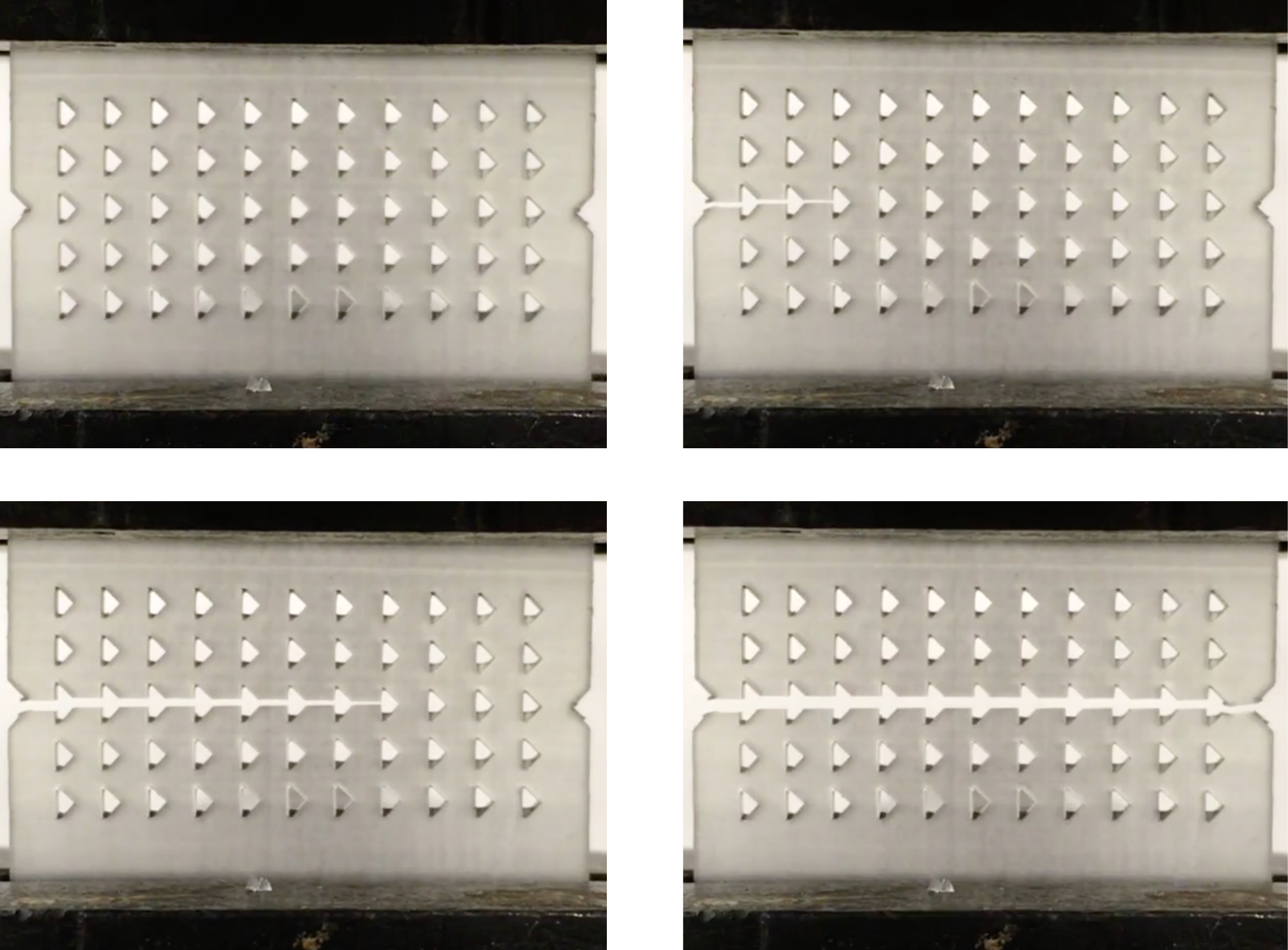}
	\vspace{-0.1in}
	\caption{Snapshots of the crack evolution in a metamaterial consisting of a two-dimensional array of voids exhibiting directional asymmetry of its effective toughness.}
\label{fig:array}
\end{figure}

The asymmetry of fracture toughness is demonstrated in a metamaterial consisting of a two-dimensional array of voids shown in Figure~\ref{fig:array} and loaded in uniaxial tension.  Even though the specimen and loading are symmetric  on a scale large compared to that of individual voids (microscale), the crack propagates from left to right indicating that the effective toughness is smaller in one direction compared to that in the other.  

To understand this, we numerically compute the effective toughness of the metamaterial following Hossain {\it et al.}~\cite{Hossain-2014}.  We take a region large compared to the microstructure and rip it apart at a constant macroscopic rate by applying a surfing boundary condition.  This is a steadily translating opening displacement ${\underline u}(x,y,t) = {\underline U}(x - Vt,y)$ imposed on  the boundary.  Here, we take  $U$ to be the displacement of a mode-I crack in plane-stress state~\cite{Zehnder-2012}
with steady velocity $V=1$ (See Figure S1 in Supplemental Information (SI)).
We compute  the fracture process using a phase field method (see Methods) with no restrictions on the crack set: pinning, kinking, branching, distal nucleation are all allowed.  We compute the macroscopic driving force on the boundary using the $J-$integral \cite{Rice-1968}.  While  the  $J-$integral can be path dependent at the scale of the heterogeneities, it reaches an asymptotic limit for large paths distal from the crack tip; further, this limit is
path-independent on paths that are sufficiently large compared to the underlying microstructure and far away from the crack  \cite{hb_jam_16}.  The driving force  ($J$) fluctuates as the fracture process negotiates the microstructure, but eventually reaches a steady pattern.  The effective toughness is the maximum of this steady pattern $J(t)$ since this is the smallest driving force necessary to drive the crack set macroscopically.   The approach has been extensively tested \cite{Hossain-2014} and experimentally verified  \cite{hetal_jmps_18,Brodnik-Hsueh-EtAl-2020a} (See Figure S2 in SI).   Importantly, the effective toughness depends only on the material and overall direction and is independent of $U$ and $V$.

\begin{figure}
\centering
\includegraphics[width=0.95\columnwidth]{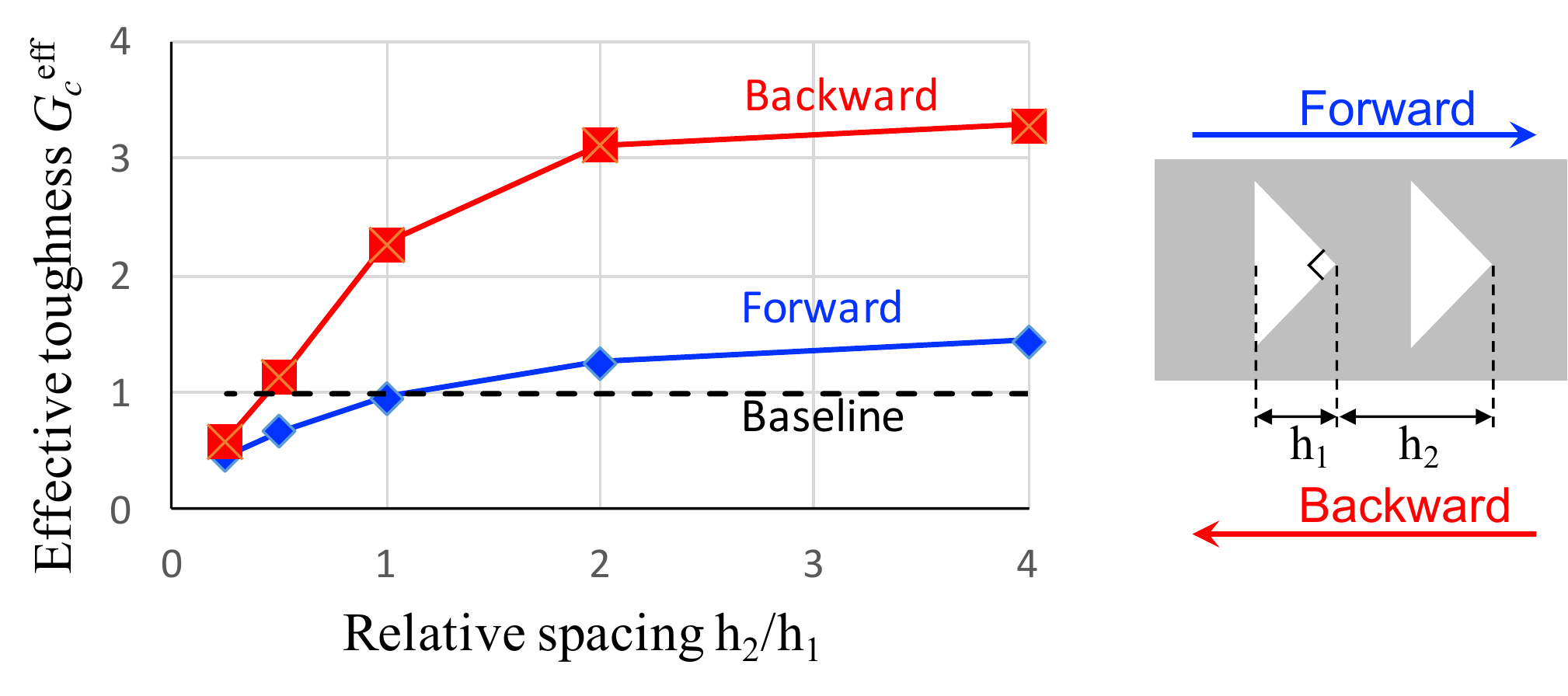}
\vspace{-0.2in}
\caption{Asymmetry in toughness in a metamaterial consisting of a two-dimensional array of voids computed using the surfing boundary conditions.  Effective toughness is normalized by the value of the base material.
\label{fig:surf}}
\end{figure}

The computed effective toughness of a metamaterial of the type shown in Figure~\ref{fig:array} is shown in Figure~\ref{fig:surf} for various cases (See SI Figure S3 for details).  Briefly, the crack propagates intermittantly at the microscale: it is pinned at each inclusion until a higher applied driving force unpins it, following which it jumps to the next inclusion.   The effective toughness in the forward direction is significantly lower than that in the backward direction resulting in the asymmetry of  toughness.  Importantly, both values  are  larger than that of the base material.   Thus, \emph{the asymmetry of toughness is not achieved by embrittling the material in one direction, but rather by asymmetrically toughening the material in both directions}.  

It is known from the study of layered materials that cracks are pinned and have to renucleate at compliant-to-stiff interfaces~\cite{hh_ijss_89,hetal_jmps_18}, but not by stiff-to-compliant interfaces.  So, at the microscale, the crack can easily enter the inclusion but has difficulty exiting it.  In the forward direction, it sees a notch where it can renucleate relatively easily.  However, in the backward direction, it sees a flat interface that it has difficulty penetrating.  This causes asymmetry while retaining superior toughness in both directions.  Designed properly, the asymmetry or difference in effective toughness can be about twice as large as the toughness of the original medium.
Finally, the effective toughness in both directions depends on length-scales or spacing, and this is well understood~\cite{Hossain-2014}.
Briefly, at very small spacing (smaller than the length-scale of the so-called macroscopic $K$-dominant zone where the macroscopic crack-tip senses and explores the stress field), the crack sees a homogeneous medium and is not pinned. 
The amount of pinning, and therefore, the effective properties increase with spacing before eventually saturating.

\begin{figure}[t]
\centering
\includegraphics[width=0.95\columnwidth]{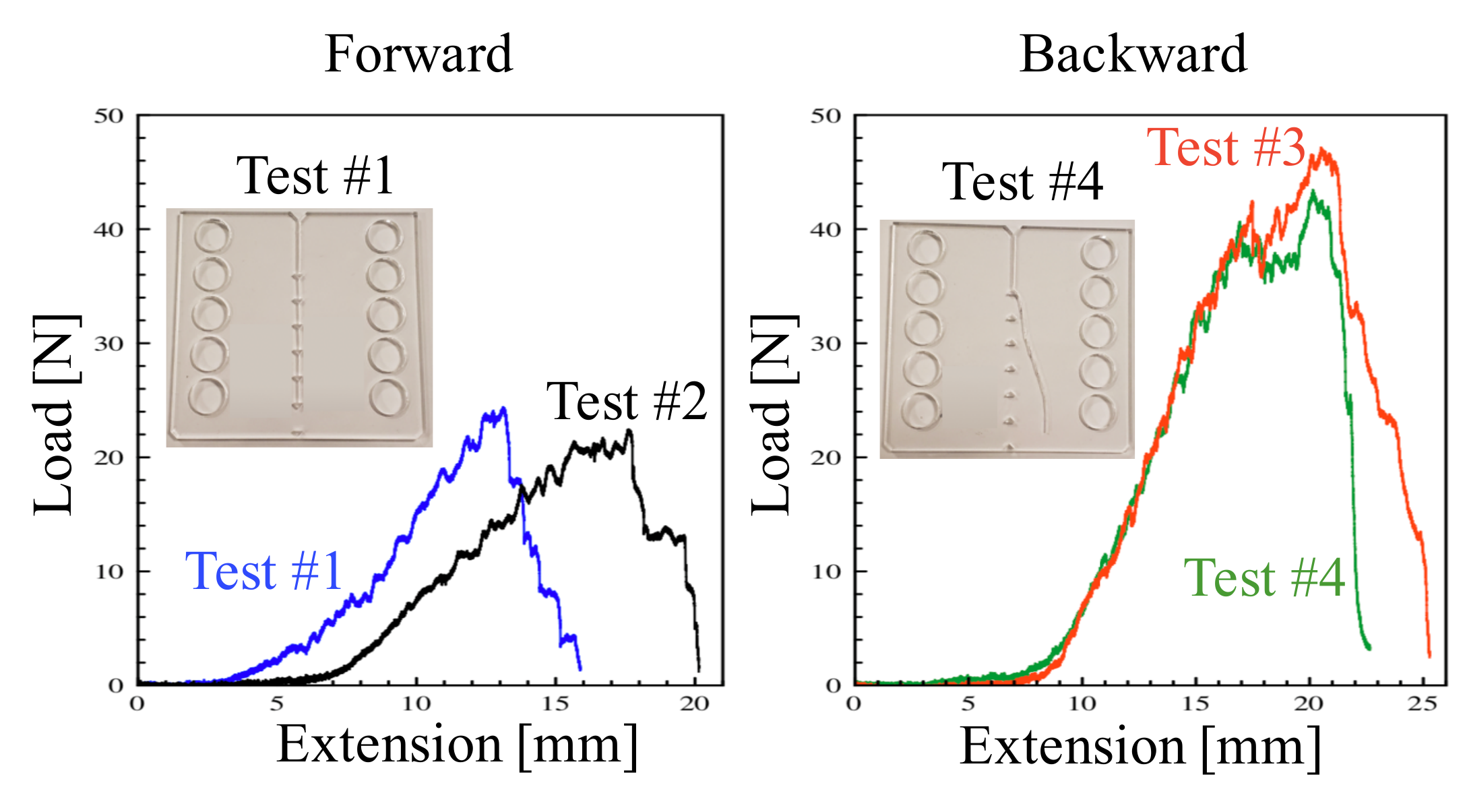}
\vspace{-0.2in}
\caption{Asymmetry in toughness in PMMA specimens tested on a rail.  
The insets show failed specimens. \label{fig:rail}}
\end{figure}

To test this idea, specimens of poly-methyl-methacrylate (PMMA) with a row of triangular voids were loaded on a rail where the loading device seeks to rip the material apart from one end as in the surfing boundary conditions  (see Methods and SI Figure S4 for details).  It is convenient to work with a specimen with a single row of  voids instead of a metamaterial for experimental reasons, but we have verified numerically that the former is representative of the latter.
The load-extension curves in both the forward and backward directions are shown in Figure~\ref{fig:rail}.  
In the forward direction, the force increases steadily till it reaches a critical value at which point a crack nucleates at the notch and rapidly advances to the first inclusion where it is pinned.  
Subsequently, the crack propagates in an intermittent manner being successively pinned and advancing rapidly to the next inclusion.  Each jump is accompanied by a load drop and each pinning phase by a load increase. 
The propagation in the backward direction is markedly different: the initial crack is nucleated as before but it is strongly pinned by the first inclusion and the load increases to almost twice the value required in the forward direction.  
At this point, the sample fails catastrophically as a crack nucleates at one of the corners of the inclusion.  
While the peak load is always higher, the crack path in the backward direction may vary (and is sensitive to the alignment in the loading device). Nonetheless, these data confirm the fracture diode concept, in which the favored fracture direction in this metamaterial design is in the forward orientation.

\begin{figure}[t]
\centering
\includegraphics[width=\columnwidth]{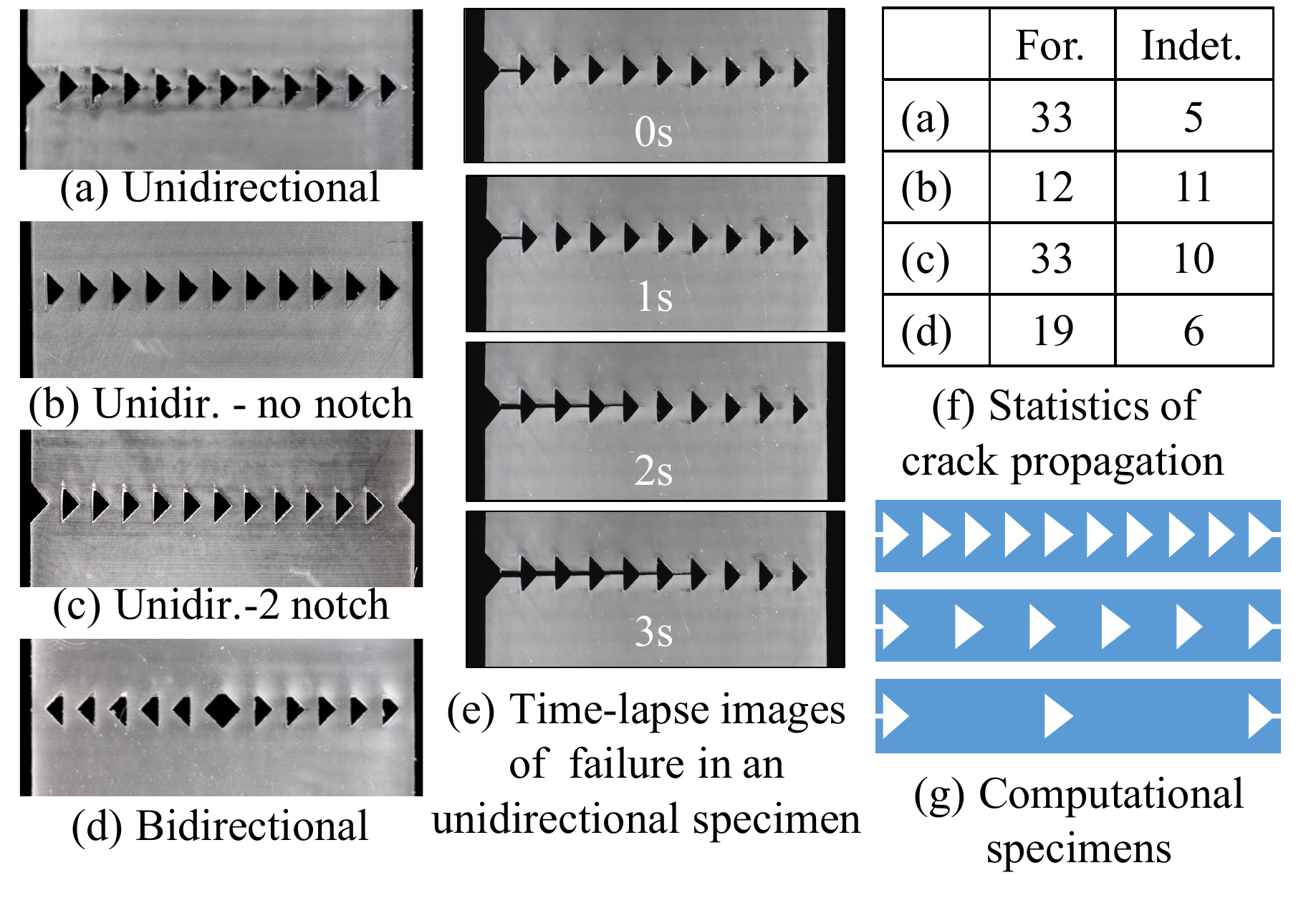}
\vspace{-0.3in}
\caption{Asymmetry in toughness in 3D printed specimens tested in uniaxial tension tests.  
(a-d) Specimen geometries.  (e) Time lapse sequence of images of a test on a uniaxial specimen (The clock begins at crack initiation).  
(f) Table of observed crack propagation direction for all tests performed (For. = Forward, Indet=Indeterminate).
(g) Specimen geometries used for computations.  Images of all specimens (a-e,g) are cropped to show the operative section. 
\label{fig:uniaxial}}
\end{figure}

The asymmetry is further established by subjecting a  series of metamaterial designs to uniaxial tension tests as in Figure~\ref{fig:array}.  Again, we focus on a single row of voids though it is representative of the metamaterial.
A centro-symmetric  design would fail with a crack propagating in either direction, but an asymmetric specimen would only fail with the crack propagating in the forward direction.  
Four designs were 3D printed with an array of triangular inclusions as shown in Figure~\ref{fig:uniaxial} and tested in uniaxial tension (see Methods and SI Figure S5 for details).   
Figure~\ref{fig:uniaxial}(e)  shows 
a series of time lapse images of the crack propagating in the forward direction.  
The figure also shows the statistics of failure: the vast majority of specimens failed with the crack propagating exclusively in the forward direction.  
Further, fractography (SI Figure S6) indicates that even the local propagation is in the forward direction; the crack nucleates at the tip of the inclusion and propagates locally in the forward direction to the next inclusion.  
In fact, even in specimens that did not completely fail in the forward direction, local failure  occurred in the forward direction.  While these results show that the microstructure does generally `rectify' the crack propagation direction, it does not always do so.

In other words, true fracture rectification behavior is somewhat subtle.   

To understand this, we study the state of stress in three computational examples shown in Figure \ref{fig:uniaxial}(g) under uniaxial tension.  The specimens have the same width, but have different number of inclusions (Details in SI Figure S7).  The first and the last inclusions are at the same location relative to the edge so that the spacing between inclusions change with the number.  If the first and last inclusions are close to the edge, the ligaments between the inclusions and edges break early as shown in  Figure S7(a).  As this stage, we compute the generalized stress intensity factor  (GSIF) that determines crack nucleation  \cite{tetal_jmps_18} (also SI) at the tip of each inclusion.  We find that the GSIF is higher on the first notch from the left, and roughly equal in every other notch (see SI).  However, the difference between the GSIF at the first notch and the rest increases with decreasing spacing.  In other words, as the spacing between inclusions increases further, the overall direction of crack growth becomes indeterminate; cracks nucleate at inclusion tips but the order in which the ligaments between inclusions break is  sensitive to material and manufacturing defects.  So, we need small spacing.  However, if the spacing becomes too small, the stress fields of the different inclusions overlap and the crack does not see each inclusion individually.  Thus, the overall toughness and asymmetry decreases as we saw earlier in Figure \ref{fig:surf}.  Therefore, there is an optimal spacing between the inclusions for sequential cracking.  Finally, it is useful to round out the two corners of the inclusion that are distal from the crack path.

\begin{figure}[t]
\centering
\includegraphics[width=0.95\columnwidth]{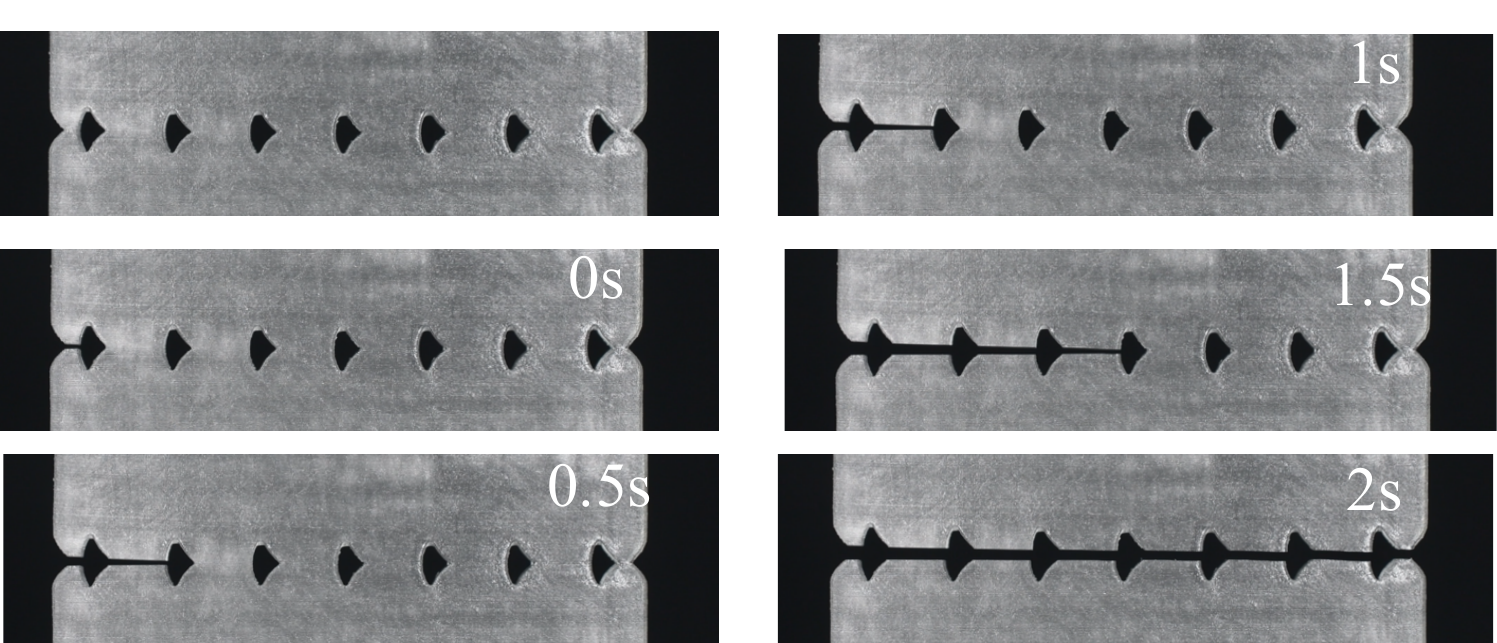}
\vspace{-0.1in}
\caption{Asymmetry in toughness in the designed fracture diode metamaterial tested in uniaxial tension tests.  The first image shows the undeformed geometry.  The inclusions have a period of 4.05mm.  The subsequent images show snapshots at 0.5 second intervals (The clock begins at crack initiation at the first inclusion).  \label{fig:final}}
\end{figure}

These principles led to the design optimized for topology with a rounded triangular void and spacing to control nucleation, shown in Figure~\ref{fig:final} (also SI Figure S8).  Twelve specimens were printed and tested.    
All twelve specimens failed with the crack propagating intermittently -- being pinned at an inclusion and then jumping to the next inclusion where it is again pinned -- in the forward direction.  
Snapshots from a representative tensile test are shown in Figure~\ref{fig:final}.  
In fact, two of the twelve specimens were pre-cracked at the opposite end with a razor blade, but this did not prevent the crack from propagating in the forward direction.
\vspace{\baselineskip}

In summary, in this letter we have established the directional asymmetry or the lack of centro-symmetry in fracture, and more broadly in bulk mechanical properties.    
We are unaware of any natural materials possessing this asymmetry.  
Importantly, the asymmetry arises from the enhancement of toughness in one direction rather than by the embrittlement in the other directions.  
The experiments presented here included a single row of voids, but the idea easily generalizes to a periodic array of 3D-printed partially filled voids~\cite{Brodnik-Hsueh-EtAl-2020a}, inclusions or other asymmetric microstructures.  Likewise, directional asymmetry would be preserved in the two- or three-dimensional metamaterials, as noted in Figure 1.
However, the design of such microstructures requires one to account for the delicate interplay between macroscopic loading, structural response, the microstructural response and the various length-scales (local and global stress field, nucleation length, etc.).

The observed asymmetry in toughness opens the way for various applications because it enables the control of crack paths and directions.    
This control can be exploited to build  resilience in structures by shielding sensitive components and function by guiding cracks away from critical regions.  
In other words, we can prescribe the failure path when failure is inevitable.  
Further, the prescription of the failure  path can enhance health-monitoring of structures.  
The control of crack paths can also enable new functionality by enabling a particular sequence of failure events without careful control of loads.

\section{Methods}
\paragraph{Variational phase-field approach to fracture}
Crack propagation in heterogeneous samples is  investigated numerically through the variational phase-field fracture approach of Bourdin {\it et al.}~\cite{Bourdin-2000, Bourdin-2008}.   
An elastic/perfectly brittle material is considered, with isotropic elastic tensor $\mathbb{C}$ (expressed in terms of the Young's modulus $E$ and Poisson's ratio $\nu$) and critical energy-release rate $G_\text{c}$, occupying a  region $\Omega\in\mathbb{R}^2$.   
In this approach, the crack is regularized on a length-scale $\ell$ by introducing a continuous (damage or fracture) field $\alpha\in[0,1]$ such that $\alpha=0$ describes the intact material and fracture is represented by regions of width proportional to $\ell$ where $\alpha$ transitions from 0 to 1. 
At each time-step, the state of the material is determined by minimizing the energy functional 
\begin{multline}
\label{eqn:energy}
\mathcal{E}_\ell(\underline{u},\alpha)=\int_\Omega \frac{(1-\alpha)^2+\eta_\ell}{2}\underline{\underline{\epsilon}}:\mathbb{C}:\underline{\underline{\epsilon}}\,d\Omega\\
+\frac{3G_\text{c}}{8}\int_\Omega\left[\frac{\alpha}{\ell}+\ell|\nabla\alpha|^2\right]\,d\Omega ,
\end{multline}
under a growth condition $\dot{\alpha}>0$ to account for the irreversible nature of the fracture process.
Above, $\underline{\underline{\epsilon}}=(\nabla\underline{u}+\nabla\underline{u}^\text{t})/2$ is the strain associated with the displacement $\underline{u}$, while $\eta_\ell$ is a small residual stiffness introduced for numerical convenience.   
A finite element discretization at mesh-size $\delta$ leads to a numerical toughness equal to $G_\text{c}^\text{num}=G_\text{c}(1+3\delta/8\ell)$ for the intact material~\cite{Bourdin-2008}.  
This approach has been shown to properly account for crack propagation, nucleation, and re-nucleation in a wide range of sitations, provided that the small parameter $\ell$ be correlated with the crack nucleation threshold~\cite{tetal_jmps_18}.
The fracture problem is solved by alternatively minimizing the total energy functional in Eq.\,\eqref{eqn:energy} with respect the two state variables $\underline{u}$ and $\alpha$.
The constrained minimization with respect to the fracture field $\alpha$ is implemented using the variational inequality solvers provided by \texttt{PETSc}~\cite{Balay-1997,petsc-web-page}, whereas the minimization with respect to displacement field $\underline{u}$ is a linear problem, solved by using a preconditioned conjugated gradient method. 
All computations are performed by means of the open source code \texttt{mef90}~\cite{mef90}. 
The equations are non-dimensionalized and the non-dimensional parameters are chosen to be $\ell=0.07$,  $\delta=0.028$, $\eta_\ell=10^{-6}$, $G_\text{c}=1$, $E=1$, $\nu=0.2$.

\paragraph{Specimen fabrication}
The PMMA specimens for the tests in Figure~\ref{fig:rail} were fabricated from a 3.175mm thick sheet using a Universal ILS9 (Tech-Labs, Katy, TX) laser cutter.   
The specimen geometry is shown in the SI.  
The 3D printed specimens were printed using digital light processing (DLP) printing on an Autodesk Ember 3D Printer (Autodesk, San Rafael, CA).  
Samples were made from commercially available Standard Clear PR48 printing resin, a urethane acrylate photopolymer.   
For the uniaxial specimens in Figure~\ref{fig:uniaxial}, the gauge length is \SI{60}{\milli\meter} and the triangle has a base of length \SI{3}{\milli\meter} and spacings of \SIlist{0.5;1;1.5}{\milli\meter}.  
The dimensions of the fracture diode in Figure~\ref{fig:final} are provided  in SI.

\paragraph{Mechanical Testing}
 We use two different modes of mechanical testing.  The first is conventional uniaxial loading where a rectangular specimen is gripped along two edges and a uniform displacement is applied across each edge.   These were performed on an Instron 5892 load frame (Instron, Norwood, MA)  at a constant displacement rate of \SI{1}{\milli\meter\per\minute} and replicates of each sample type were rotated and mirrored randomly to ensure that no bias was introduced due to the innate directionality of the DLP printing process.   
For each test, the load and displacement were recorded using data from the load cell and the failure behavior of the sample itself was recorded with a Nikon D7500 (Nikon, Tokyo, Japan) digital camera at a rate of 30 frames per second.  
Loading data and video were synchronized through visible failure events.  
After testing, video recordings of failure were then reviewed frame-by-frame using the post-production film software DaVinci Resolve (Blackmagic Design, Port Melbourne, Australia) to classify failure based on criteria of forward or indeterminate failure based on behavior predicted by analogous simulations

The second mode of mechanical testing is an unconventional method that seeks to rip a specimen apart from one end using a rail following Hsueh {\it et al.}~\cite{hetal_jmps_18}  (see SI).  The rectangular specimen contains a row of circular holes near two opposing edges.  A bushing passes through each hole, and the bushings are guided along a wedge shaped rail system so that pairs of opposing holes are pulled apart sequentially.  The wedge-shaped rail has an angle of  \SI{2.2}{\degree}) and is loaded using an Instron 5892 load frame at a constant displacement rate of \SI{6}{\milli\meter\per\minute}.

\vspace{1em}
This work was conducted while Brodnik and Brach were at Caltech.
We are grateful to Paolo Celli and Kevin Korner for their help in preparing the PMMA specimens.  
We gratefully acknowledge the financial support of the U.S. National Science Foundation (Grant No. DMS-1535083 and 1535076) under the Designing Materials to Revolutionize and Engineer our Future (DMREF) Program. 
The development of the numerical codes used here was supported in part by the U.S. National Science Foundation  DMS-1716763.
The numerical simulations were performed at the Caltech high performance cluster supported in part by the Moore Foundation.

\bibliographystyle{apsrev4-1}
\bibliography{mainb.bib}

\end{document}